\documentclass[referee]{aa}
\usepackage{txfonts}
\usepackage{graphicx}
\usepackage{setspace}
\usepackage{natbib}
\usepackage{longtable}
\usepackage{amssymb}
\bibpunct{(}{)}{;}{a}{}{,}

\begin{document}

\newcommand{\fe}{[\ion{Fe}{ii}]}
\newcommand{\Ti}{[\ion{Ti}{ii}]}
\newcommand{\s}{[\ion{S}{ii}]}
\newcommand{\oi}{[\ion{O}{i}]}
\newcommand{\pii}{[\ion{P}{ii}]}
\newcommand{\Ni}{[\ion{Ni}{ii}]}
\newcommand{\brg}{Br$\gamma$}
\newcommand{\mdot}{$\dot{M}_{jet}$}
\newcommand{\mjet}{$\dot{M}_{jet}$}
\newcommand{\mh}{$\dot{M}_{H_2}$}
\newcommand{\Ne}{n$_e$}
\newcommand{\h}{H$_2$}
\newcommand{\kms}{km\,s$^{-1}$}
\newcommand{\um}{$\mu$m}
\newcommand{\lam}{$\lambda$}
\newcommand{\msyr}{M$_{\odot}$\,yr$^{-1}$}
\hyphenation{mo-le-cu-lar pre-vious e-vi-den-ce di-ffe-rent pa-ra-me-ters ex-ten-ding a-vai-la-ble}

\title{IR diagnostics of embedded jets: kinematics and physical characteristics of the HH46-47 jet \thanks{Based on observations collected at the European Southern 
Observatory, Paranal, Chile (ESO programmes 0.74.C-0286(A)).}}
\author{Rebeca Garcia Lopez \inst{1,2} \and Brunella Nisini \inst{1} \and Jochen Eisl\"offel \inst{2} \and Teresa Giannini \inst{1} \and Francesca Bacciotti \inst{3} \and Linda Podio \inst{4} }

\offprints{R. Garcia Lopez, \email{garcia@oa-roma.inaf.it}}

\institute{INAF-Osservatorio Astronomico di Roma, Via di Frascati 33, I-00040 Monteporzio Catone,
Italy \and Th\"uringer Landessternwarte Tautenburg, Sternwarte 5, D-07778 Tautenburg, Germany \and 
INAF-Osservatorio Astrofisico di Arcetri, Largo E. Fermi 5, I-50125 Florence, Italy \and Dublin Institute for Advanced Studies, 31 Fitzwilliam Place, Dublin 2, Ireland}

%
\date{Received date; Accepted date}
%
%
%
\titlerunning{Velocity resolved observations of HH46-47}
\authorrunning{Garcia Lopez, R. et al.}

\abstract
{We present an analysis of the kinematics and physical properties of the Class I driven jet HH46-47 based on IR medium and low resolution spectroscopy obtained with ISAAC on VLT. }
{Our aim is to study the gas physics as a function of the velocity and distance from the source and compare the results with similar studies performed on other Class I and classical T Tauri jets, as well as with existing models for the jet formation and excitation.}
{The ratios and luminosities of several important diagnostic lines (e.g. \fe\,1.644, 1.600\,\um, \pii\,1.189\,\um, and H$_2$ lines) have been used  to derive  physical parameters such as electron density, H$_2$ temperature, iron gas-phase abundance and mass flux. \fe\,1.644\,\um\ and H$_2$\,2.122\,\um\ Position Velocity Diagrams (PVDs) have been in addition constructed  to study the kinematics of both the atomic and molecular gas.}
{Within 1000-2000\,AU from the source the atomic gas presents a wide range of radial velocities, from $\sim$\,-230\,\kms\ to $\sim$\,100\,\kms. Only the gas component at the highest velocity (High Velocity Component, HVC) survives at large distances. The H$_2$, shows, close to the source, only a single velocity component at almost zero velocity, while it reaches higer velocities (up to $\sim$95\,\kms) further downstream. Electron densities (n$_e$) and mass ejection fluxes (\mdot) have been measured separately for the HVC and for the component at lower velocity (LVC) from the \fe\ lines. n$_e$ increases with decreasing velocities with an average value of $\sim6000$\,cm$^{-3}$ for the LVC and $\sim4000$\,cm$^{-3}$ for the HVC, while the opposite occurs for \mdot\ which is  $\sim0.5-2\times10^{-7}$\,M$_{\odot}$yr$^{-1}$ and $\sim0.5-3.6\times10^{-8}$\,M$_{\odot}$yr$^{-1}$ for the HVC and  LVC, respectively. The mass flux carried out by the molecular component, measured from the H$_2$ lines flux, is $\sim4\times10^{-9}$\,M$_{\odot}$yr$^{-1}$.
We have estimated that the Fe gas phase abundance is significantly lower than the solar value, with $\sim$ 88\% of iron still depleted onto dust grains in the internal jet region. This fraction decreases to $\sim$58\%, in the external knots.}
{Many of the derived properties of the HH46-47 jet are common to jets from YSOs in different evolutionary states. The derived densities and mass flux values are typical of Class I objects or very active T Tauri stars. However, the spatial extent of the LVC and the velocity dependence of the electron density have been so far observed only in another Class I jet, the HH34 jet, and  are not explained by the current models of jet launching.}


\maketitle
\maketitle
%

\section{Introduction}

Protostellar jets are tightly associated with the first stage of protostar evolution. They are believed to remove angular momentum, disrupt infalling material from the cloud, inject turbulence in the ISM and modify its chemistry. Little is known, however, on how protostellar jets originate. There are two principal schools of thought regarding where protostellar jets are launched from: the X-wind and the Disc-wind models (\citealt{shu95, ferreira97}). While disc-wind models assume ejection from a large spread of radii from the disc, X-wind models assume that the ejection of material takes place only at a single annulus at the interaction region between the stellar magnetosphere and the inner edge of the disc. Both models were born in an attempt to explain the kinematical and physical properties observed at optical wavelengths in jets from classical TTauri stars (e.g., \citealt{bacciotti00,lavalley97}). Very few studies have been done, however, regarding the physical properties of Class I jets. Since these objects are still highly embedded in their molecular clouds, they are affected by a large extinction. Thus, it is not possible to detect the region near to the exciting source at optical wavelengths. In this context, infrared spectroscopy is a very useful tool to probe the physical properties and kinematics of Class I jets close to their base, where their characteristics have not yet been modified by the interaction with the surrounding material.
\begin{figure}
\resizebox{\hsize}{!}{\includegraphics{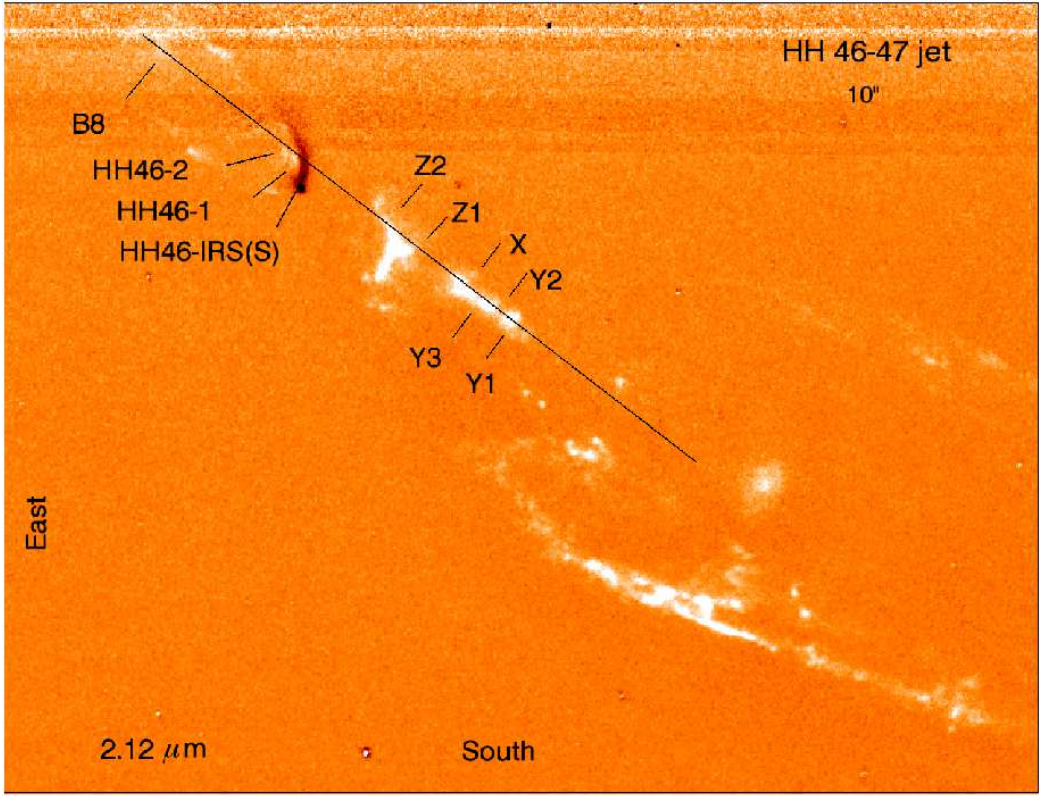}}
\caption{Position of the ISAAC slit superimposed on the H$_2$ 2.12\,\um\, image of the HH46-47 jet. At the source (HH46-IRS) and knot S position  a negative residual from the continuum substraction is present. Individual knots along the jet are identified following and our own (see text) nomenclature and that of \cite{jochen94_hh46}. The northeastern knots correspond to the blue-shifted jet.}
  \label{fig:imaging}
\end{figure}

Several studies have been carried out in recent years employing near-IR (NIR) line diagnostic on protostellar jets (e.g., \citealt{nisini_hh1, linda06, davisSVS13,takami}). The main lines used to probe physical and kinematical properties in NIR are \fe\ (e.g. 1.644\,\um, 1.600\,\um) and \h\ lines (e.g. 2.122\,\um). These studies revealed that the kinematical structure at the jet base of Class I and classical T Tauri stars (CTTSs) jets is very similar. Both show a broad line emission near the source consisting of a collimated and large scale jet at high velocity (the so-called HVC) and a slower velocity component (the so-called LVC) that is only detected around the central source. Noticeably, this kind of velocity structure is often observed in both the atomic and molecular components of the jets, the latter probed through the H$_2$ emission and called Molecular Hydrogen Emission Line regions \citep{davis_MHEL}. Concerning the physical properties, Class I jets show higher densities and mass loss rates than CTTS jets, as expected for less evolved objects.
Both disc-wind and X-wind models predict the presence of a large spread of velocity at the jet base as shown by few synthetic position velocity diagrams (PVDs) constructed for both models (\citealt{shang98,pesenti04}), although attempts to reproduce the specific characteristics (e.g. kinematics, presence of the MHEL region, physical parameters) displayed by the Class I jets are still lacking . 
We have now started a project aimed to determine few basic physical parameters of embedded jets as a function of both velocity and distance from the driving source, using near-IR (NIR) medium resolution (R$\sim$10\,000) spectra taken with ISAAC on the Very Large Telescope (VLT). The goals of this study are to make a quantitative comparison between  the physical properties of Class I and CTTS jets and  to constrain the origin of the different velocity components and the mechanism responsible for their excitation. 
In a previous paper (\citealt{rebeca08}), we have applied our NIR diagnostic techniques to two jets, HH34 and HH1. 


\begin{table*}
\begin{minipage}[t]{\textwidth}
\caption{Lines observed in the knot Z1 of the HH46-47 jet.                                                                               }
\label{tab:z1_lines}
\vspace{0.1cm}
\centering
\renewcommand{\footnoterule}{}  
\begin{tabular}{cccccccc}
\hline     \hline 
Line id. & $\lambda$ &  F & $\Delta$F & Line id. & $\lambda$ &  F & $\Delta$F\\
 ~       &  (\um)    & \multicolumn{2}{c}{(10$^{-16}$ erg cm$^{-2}$ s$^{-1}$)} &~        & (\um)    & \multicolumn{2}{c}{(10$^{-16}$ erg cm$^{-2}$ s$^{-1}$)} \\
\hline 
\Ti\,$^2$G$_{7/2}$-$^4$F$_{7/2}$    & 1.140 & 8.3 & 3.5 & 	H$_2$ 1-0 S(6)			   & 1.788 & 9.4 & 0.5 \\
H$_2$ 2-0 S(1)   		    & 1.162 & 5.8 & 2.2 & 	\fe\,a$^4$D$_{3/2}$-a$^4$F$_{3/2}$ & 1.797 & 9.8 & 1.1\\
\pii\,$^3$P$_2$-$^1$D$_2$	    & 1.188 & 3.5 & 1.0 & 	\fe\,a$^4$D$_{5/2}$-a$^4$F$_{5/2}$ & 1.800 & 7.1 & 0.8\\
\fe\,a$^4$D$_{7/2}$-a$^6$D$_{9/2}$ & 1.257 & 106.0 & 0.8& 	\fe\,a$^4$D$_{7/2}$-a$^4$F$_{7/2}$ & 1.810 & 38.2 & 2.1\\
\fe\,a$^4$D$_{1/2}$-a$^6$D$_{1/2}$ & 1.271 & 5.9 & 1.0  & 	H$_2$ 2-1 S(5)			   & 1.945 & 11.0 & 1.5  \\
\fe\,a$^4$D$_{3/2}$-a$^6$D$_{3/2}$ & 1.279 & 7.8 & 0.1  & 	H$_2$ 2-1 S(4)			   & 2.004 & 2.7 & 0.3 \\
\fe\,a$^4$D$_{5/2}$-a$^6$D$_{5/2}$ & 1.295 & 11.9 & 0.9 & 	H$_2$ 1-0 S(2)			   & 2.034 & 26.1 & 0.6 \\
\fe\,a$^4$D$_{3/2}$-a$^6$D$_{1/2}$ & 1.298 & 6.7 & 0.1 & 	H$_2$ 2-1 S(3)			   & 2.074 & 8.1 & 0.5  \\
\fe\,a$^4$D$_{7/2}$-a$^6$D$_{7/2}$ & 1.321 & 306.0 & 0.8  & 	H$_2$ 1-0 S(1)			   & 2.122 & 68.5 & 0.3 \\
\fe\,a$^4$D$_{5/2}$-a$^6$D$_{3/2}$ & 1.328 & 8.4 & 0.9  & 	H$_2$ 2-1 S(2)			   & 2.154 & 3.9 & 0.3	 \\
\fe\,a$^4$D$_{7/2}$-a$^6$D$_{5/2}$ & 1.372 & 71.7 & 1.0 & 	H$_2$ 2-1 S(1)		           & 2.248 & 8.5 & 0.5 \\
\fe\,a$^4$D$_{5/2}$-a$^4$F$_{9/2}$ & 1.534 & 173.0 & 0.7 & 	H$_2$ 3-2 S(1)			   & 2.386 & 1.6 & 0.4\\
\fe\,a$^4$D$_{3/2}$-a$^4$F$_{7/2}$ & 1.600 & 10.6 & 0.5 &	H$_2$ 1-0 Q(1)			   & 2.407 & 50.0 & 0.8     \\ 
\fe\,a$^4$D$_{7/2}$-a$^4$F$_{9/2}$ & 1.644 & 160.0 & 0.6 &  	H$_2$ 1-0 Q(2)		           & 2.413 & 18.3 & 0.9 	\\ 
\fe\,a$^4$D$_{1/2}$-a$^4$F$_{5/2}$ & 1.664 & 7.0 & 0.7 & 	H$_2$ 1-0 Q(3)			   & 2.424 & 47.4 & 0.9  \\
\fe\,a$^4$D$_{5/2}$-a$^4$F$_{7/2}$ & 1.677 & 142.0 & 0.3 & 	H$_2$ 1-0 Q(4)	                   & 2.437 & 18.9 & 1.3 \\
H$_2$ 1-0 S(9)			   & 1.687 & 4.2 & 0.5 & 	H$_2$ 1-0 Q(5)			   & 2.455 & 42.9 & 1.5    \\
\fe\,a$^4$D$_{3/2}$-a$^4$F$_{5/2}$ & 1.712 & 2.1 & 0.4 &  	H$_2$ 1-0 Q(5)			   & 2.455 & 42.9 & 1.5    \\
\fe\,a$^4$D$_{1/2}$-a$^4$F$_{3/2}^*$\footnotetext{$^*$Blend of the two indicated lines.} 
				   & 1.745 & 6.3 & 0.8 &	H$_2$ 1-0 Q(7)		  	 & 2.500 & 47.4 & 3.7  \\  
H$_2$ 1-0 S(7)$^*$       	   & 1.748 & 150.0 & 0.6&					&	&	&	\\

\hline
\end{tabular}
\end{minipage}
\end{table*}

We now extend this previous study to the HH46-47 jet. This known Class I jet is at a distance of $\sim$450\,pc emerging from a dense reflecting nebula. The jet  is located at the border of a Bok globule and powers an atomic and molecular bipolar flow. The driving source (HH46-IRS) is in a phase of high accretion (\citealt{simone08}) and is known to be a deeply embedded binary system (\citealt{reipurth00}). The northeastern (blue-shifted) portion of the flow is detected at optical wavelengths giving rise to strong HH objects \citep{hartigan93,hartigan05}. Recently, the kinematics of the internal region of this optical jet has been studied in details by \cite{nishikawa08}.
At infrared wavelengths, the HH46 jet and its red-shifted counter-jet are detected down to the central source (\citealt{jochen94_hh46IR}), thus this object represents a very suitable target for our NIR spectroscopy study.

In Sect. 2, we present the observations and the data reduction. In Sect.\,3 we describe the results of our observations, while the kinematics of the atomic and molecular components are presented  in Sect.\,4. In Sect.\,5 and 6 we describe the spectral diagnostic analysis applied to ISAAC spectra obtained at low and medium resolution, respectively.
We have compared the results found on the HH46-47 jet with the HH34 jet in Sect.\,7 and  the main conclusions of our work are finally summarised in Sect.\,8. 
\begin{figure*}[!ht]
\resizebox{\hsize}{!}{\includegraphics{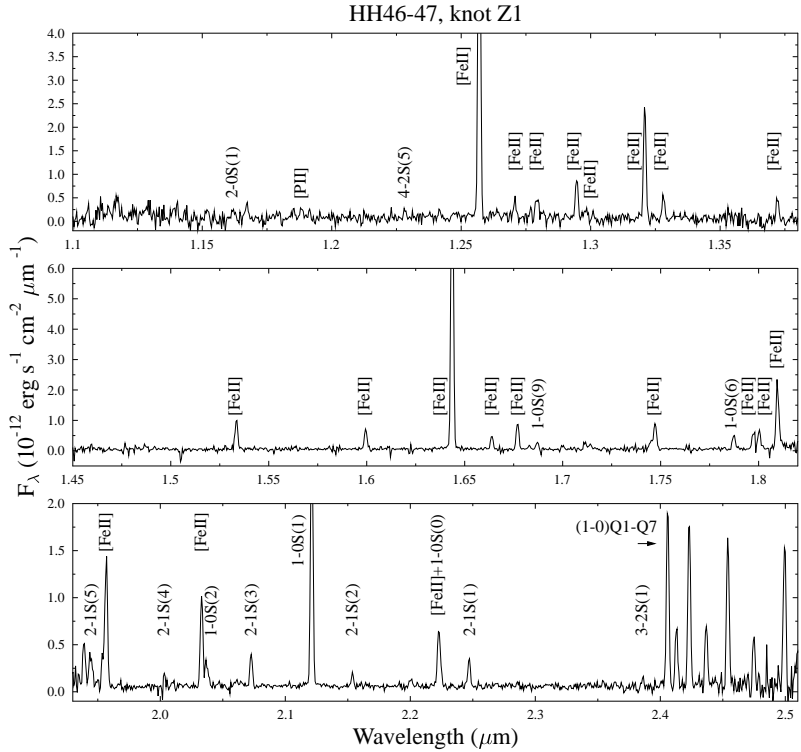}}
\caption{LR spectrum from 1.00 to 2.50\,\um\, of the knot Z1 in the HH46-47 jet. The stronger lines are identified.}
\label{fig:knot_z1}
\end{figure*}
\section{Observations and data reduction}

The observations were made on 29-30 December 2004 at ESO VLT telescope on Cerro Paranal, Chile, using the infrared spectrograph and camera ISAAC. The spatial scale of the instrument is 0\farcs\,148/pixel. We have taken low resolution spectra in the J-band (1.1-1.4\,\um\,) and both, medium and low resolution spectra in the H (1.42-1.82\,\um) and K (1.82-2.5\,\um) bands. The adopted slit width were 0\farcs6 and 0\farcs3 for the low and medium resolution, respectively, corresponding  to nominal resolutions of 860 (in J), 840 (in H) and 750 (in K) 
at low resolution (LR) and 10\,000 (in H) and 8900 (in K) at medium resolution (MR). 
The slits were all aligned along the jet, with a position angle (P.A.) of $\sim$57\degr.
Total integration times were 600\,s for each filter in the LR mode and, 3600\,s and 7200\,s for the K- and H-band in the MR mode.

Data reduction was performed using standard IRAF\footnote{IRAF (Image Reduction and Analysis Facility) is distributed by the National Optical Astronomy Observatories, which are operated by AURA, Inc., cooperative agreement with the National Science Foundation.} tasks. Wavelength calibration was done using the atmospheric OH emission at  MR and an argon lamp at LR. The atmospheric spectral response was corrected by dividing the object spectra by the spectrum of a telluric standard star for both modes. Flux calibrations were performed using photometric standards observed at an air mass and seeing similar to our target observations. This procedure probably leads to some flux losses due to the finite slit width with respect to the jet width, especially in the case of the bow shaped knot Z1. In MR spectroscopy, the continuum of the HH46-IRS source was subtracted using the IRAF task BACKGROUND. 

The reduced MR spectra on the source position have been already presented in \cite{simone08} (reduced and analysed in the same way as explained above) and thus they will not be discussed in the present paper.
\section{Results}

\begin{figure*}
\includegraphics[totalheight=10cm]{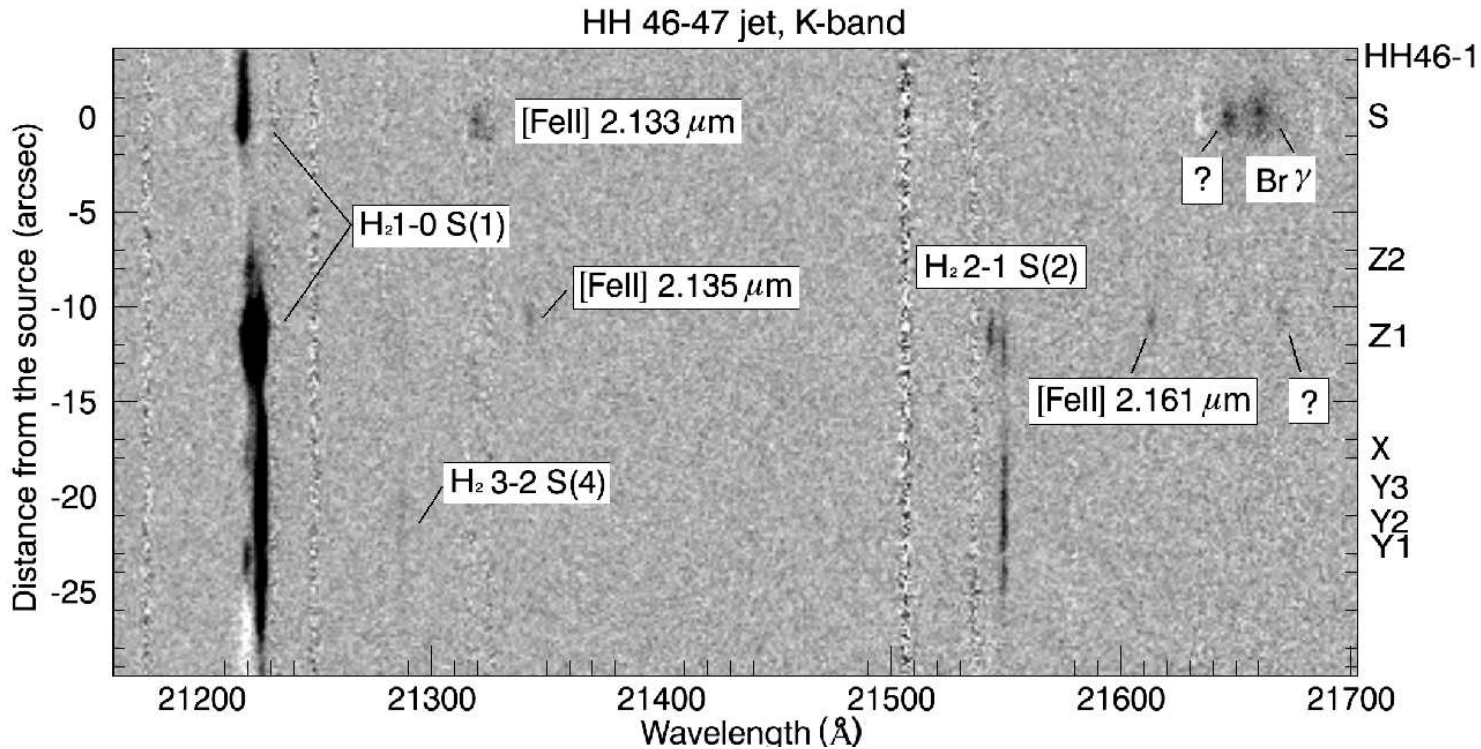}
\caption{MR continuum subtracted spectral image of the HH46-47 jet in the K-band. The distance from the source is represented on the left-hand Y-axis, while the knots are labelled on the right-hand.}
\label{fig:K-band}
\end{figure*}

Fig.\,\ref{fig:imaging} shows a H$_2$\,2.12\,\um\ continuum-subtracted image of the HH46-47 jet. The slit position is depicted over the image and the covered knots are named following the nomenclature of \cite{jochen94_hh46}. 
In addition, we name as knot S the emission that peaks at the source position and extends for $\sim$3\arcsec\ in the blue-shifted jet.
The observations cover the blue-shifted jet  knots HH46-1 and 2 and the knot B8, which is located at the apex of a bow shock. In the red-shifted lobe, the slit comprises knots Z and Y in \cite{jochen94_hh46} that have been here separated in knots Z1 and Z2, and Y1, Y2 and Y3. 

The low resolution spectra present several lines of both atomic and molecular species in the different knots. Figure\,\ref{fig:knot_z1} shows the complete spectrum of knot Z1, which is among the brightest ones along the jet. A list of the observed lines, fluxes and identification in knot Z1 is also reported in Table\,\ref{tab:z1_lines}, while Tables\,\ref{tab:atomic_fluxes} and \ref{tab:molecular_fluxes} in the Appendix\,\ref{app:fluxes} list all the lines detected in the other knots.  \fe\ transitions from the first 12 fine-structure  levels, and \h\ transitions  from the two first vibrational levels have been detected in all of the knots at various intensity levels. The knot S is, together with the knot Z1, the richest in ionic lines. Here \pii\ $^3P_2-^1D_2$\,1.189\,\um  has also been detected.

The medium resolution K-band spectra show several lines, the brightest being  \h\,1-0S(1), 3-2S(4) and 2-1S(2),  \fe\,2.1349 and 2.1609\,\um\, and Br$\gamma$ (see, Fig.\,\ref{fig:K-band}). 
In addition, the  \fe\,1.600 and 1.644\,\um\ lines have been detected in the H-band spectra. The presence of the Br$\gamma$ near the source and the \h\,3-2S(4) line at a distance of 10\arcsec\ to 25\arcsec\ from HH46-IRS testifies to the high excitation conditions in these regions. In addition, the on-source spectrum shows other features, as the Br14 and Br12 and CO transitions (see also, \citealt{simone08}).

\section{Kinematics}

\begin{figure}[!ht]
\begin{center}
\includegraphics[totalheight=18cm]{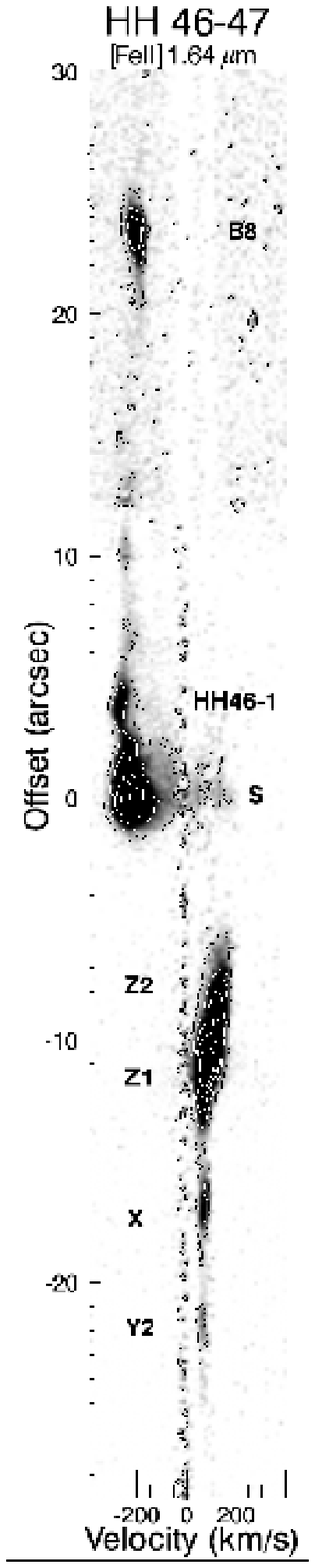}
\includegraphics[totalheight=18cm]{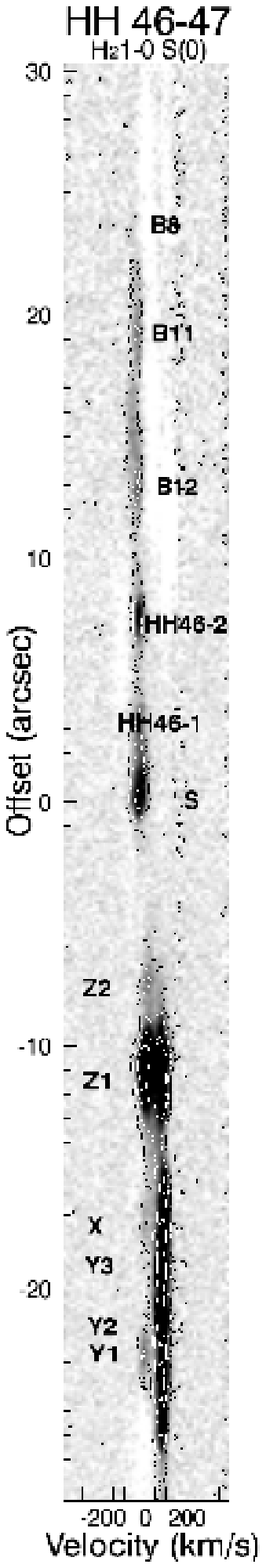}
\caption{\label{fig:PVDs}
Continuum-subtracted PVDs for the \fe\,1.644\,\um\ and \h\,2.122\,\um\, lines along the HH46-47 jet. Contours show values of 3, 9, \ldots\,, 243\,$\sigma$ for the \fe\, line and 4.5, 13.5,\ldots,364.5\,$\sigma$ for the \h\,. On the Y axis distance from the source HH46-IRS in arcsec is reported.}
\end{center}
\end{figure}

In order to study the kinematics of the atomic and molecular component of the HH46-47 jet as a function of distance from the central source, we have constructed Position Velocity Diagrams (PVDs) of the \fe\,1.644\,\um\, and \h\,2.122\,\um\ lines (see, Fig.\,\ref{fig:PVDs}) from the MR spectral images.
The velocity has been expressed with respect to the Local Standard of Rest (LSR) and corrected for a parental cloud velocity of +20\,\kms\, (\citealt{hartigan93}). In the Y-axis the distance from the driving source HH46-IRS, in arcsec, is represented. The continuum of the source has been subtracted in both the PVDs to more clearly see the structure of the lines near HH46-IRS. The intensity scale is different in the two PVDs in order to show the different condensations in the atomic and molecular gas.

Emission down to the central source is detected for both the atomic and molecular components (knot S). The \fe\, emission broaden as it approaches the source position and spans velocities from $\sim$\,-200\,\kms\ down to 0\,\kms\ within 2\,\arcsec\ from the source. Inside 1\,\arcsec\ there is also emission at red-shifted velocities up to $\sim$\,+100\,\kms. We will call the blue-shifted component at high velocity that can be traced up to large scale the high velocity component (HVC) and the more compact emission component from $\sim$\,-150\,\kms\ to $\sim$\,0\,\kms\ the low velocity component (LVC).
These two components have been also evidenced in H$\alpha$ optical observations \citep{nishikawa08}. At variance with the ionic emission, the \h\ emission close to the source (knot S) shows only one component at nearly 0\,\kms.

The radial velocities of the \fe\,1.644\,\um\ and \h\,2.122\,\um\ lines are reported in Table\,\ref{tab:vhh46}. They have been computed by a Gaussian-fit to the line profile of every knot. The velocities of the \fe\ emission range from -235\,\kms\ to -177\,\kms\ and from +144\,\kms\ to +94\,\kms\ in the blue- and red-lobes, respectively: thus in both lobes the peak velocity decreases as the distance from the driving source increases in both lobes. The HVC and LVC in knot S peak at -213\,\kms\ and around -138\,\kms. 
At knot Z2 the emission splits in two separate velocity components (at +144\,\kms\ and +92\,\kms) whose origin is likely connected  with the presence of the large bow shock structure.

\begin{figure}[!ht]
\begin{center}
\includegraphics[totalheight=10cm]{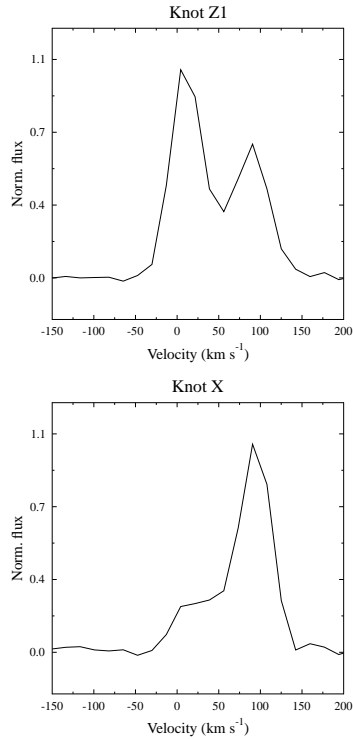}
\caption{\label{fig:h2_profiles}
Normalised line profiles of the \h\,2.122\,\um\ line for knots Z1 (upper panel) and X (lower panel). In the X-axis the LSR radial velocity  is represented. The velocities are corrected from a cloud velocity of +20\,\kms\ (\citealt{hartigan93})}
\end{center}
\end{figure}

The radial velocities of the \h\ emission vary from -15\,\kms\ to -48\,\kms\ in the blue-lobe an from +26\,\kms\ to +94\,\kms\ in the red-lobe. At variance with the \fe\ emission, knot S only shows a single velocity component at -15\kms\ while both knots Z1 and X present two velocity components at $\sim$ +90\,\kms\, and +10\,\kms.
The high and low velocity components present in these knots and the broad profiles seen in \h\ (see, Fig.\,\ref{fig:h2_profiles}) are similar to the profiles observed in other molecular jets (e.g., \citealt{davis01}). As shown in Fig.\,\ref{fig:h2_profiles}, the relative contribution of the high and low velocities components to the total flux changes from knot Z1 (where the  component at 0\,\kms\ is larger than that at high velocity) to knot X (where the component at 0\,\kms\ is significantly reduced). These profiles are similar to those expected from bow-shocks structures viewed at a certain inclination angle (\citealt{schultz05}). With respect to the \cite{schultz05} models, the HVC and LVC contribution to the total flux depends, in our case, not only on the viewing angle of the bow-shock but also on how the \h\ cooling is distributed from the apex to the wings of the bow. In particular, the profiles of the knots Z1 and X are consistent with bow-shocks seen at 30\degr\ with respect to the line of sight, but with a different relative contribution to the \h\ cooling between the apex and the wing. In knot X the apex mostly contribute to the total flux while in knot Z1 the situation is the opposite. Such an effect could be actually due to a not good alignment of the slit with the bow-shock apex in knot Z1.

\begin{table}
\begin{center}
\caption{\label{tab:vhh46}Observed radial velocities along the HH46-47 jet.}

\vspace{0.5cm}
\renewcommand{\footnoterule}{}  
    \begin{tabular}[h]{c|c|cc}
 \hline     \hline \\[-5pt]

Knot & \fe\,1.644\,\um\, & H$_2$\,2.122\,\um\, \\
     &     v(\kms\,)     & v(\kms\,) \\
\hline \\[-5pt]
B8  &-177 &        \\
HH46-1  &-235&-20 \\
S &-213,(-138)& -15 \\
Z2  &+144, +92 &+26 \\
Z1  &+112 &+89, +10 \\
X &+94 &+93, +14  \\
Y3 & &+94 \\
Y2 & &+92 \\
Y1 & &+92 \\

\hline

    \end{tabular}

\end{center}

~Note: Radial velocities corrected from a cloud velocity of +20\,\kms\ (\citealt{hartigan93}) and with respect to
the LSR. The radial velocity error is 2\,\kms.
\end{table}

\section{Low resolution spectral analysis}
\subsection{Extinction}

Visual extinction values can be derived from the \fe\,1.644/1.257\,\um\ ratio: these lines share the same upper level  and thus, their theoretical ratio is a function of the frequencies and probabilities of the transition and are independent of the plasma conditions. 
To convert the observed ratios into $A_v$ values, we have adopted the radiative transition probabilities given by \cite{quinet96} using the relativistic Hartree-Fock (HFR) computer program. 
The extinction errors have been estimated directly from the flux error of each line.
The results are reported in Table\,\ref{tab:Av} for the individual knots where the S/N of the two lines is larger than 10.

\begin{table}
\begin{minipage}[t]{\columnwidth}
\caption{A$_V$ along the HH46-47 jet.}
\label{tab:Av}
\centering
\renewcommand{\footnoterule}{}  
    \begin{tabular}[h]{cc}
 \hline     \hline \\[-5pt]

Knot & 
A$_V\pm \bigtriangleup$A$_V$\footnote{Visual extinction measured from the ratio \fe\,1.64/1.25\,\um\,} \\

  & (mag)  \\
\hline \\[-5pt]

S    &  6.6$\pm$0.2 \\
Z2   &  12.5$\pm$1.6 \\
Z1   &  6.1$\pm$0.1 \\

\hline
\end{tabular}
\end{minipage}
\end{table}


A$_V$ varies from $\sim$6 in the knots Z1 and S to $\sim$12 in the knot Z2. \cite{fernandes00} found a value of A$_V=9.38\pm\,1.49$\,mag in a region that covers our knots HH46-1 and S. This value has been  however derived employing  the theoretical \fe\,1.644/1.257\,\um\ ratio given by the radiative rates  of \cite{nussbaumer88} that predict higher values with respect to the \cite{quinet96} rates (\citealt{nisini_hh1}). 
On the other hand, \cite{simone08} found an A$_V\,\sim$40\,mag on-source, indicating a large density gradient towards the central object. One should note here that the extinction found by \cite{simone08} on the basis of photometric measurements coupled with \brg\ emission, is larger than the one we found on the knot S, since \fe\ lines trace more external regions. In addition, the knot S covers a region that extends $\sim$3\arcsec\ around the source position.

The extinction in the knot Z2 is roughly twice the extinction of the knots S and Z1. This fact could be due to 
the collision of the jet with stationary material on a side of the cavity. \cite{heathcote96} note the presence of an oblique shock due to the impact of the blue-shifted lobe of the jet with the cavity at $\sim$6\arcsec\ NE of the source. The same process may occur at the same distance in the red-shifted counter-lobe, giving rise to an increase in extinction.

\begin{table}
\begin{minipage}[t]{\columnwidth}
\caption{d$v_t$,  T, N($H_2$), \mh\ along the HH46-47 jet.}
\label{tab:h2}
\centering
\renewcommand{\footnoterule}{}

    \begin{tabular}[h]{c|cccc}
 \hline     \hline \\[-5pt]

Knot & $dv_t^a$ & T$^b$ & N(H$_2$)$^c$ &  \mh$^d$ \\

     & (\kms)  & (K) & (10$^{17}$\,cm$^{-2}$) & (\msyr) \\
\hline \\[-5pt]

Z1   & 131 & $2797\pm100$ & 3.3 &  4.5$\times$10$^{-9}$\\
X    & 19 & $2078\pm42$  & 3.7 &  7.3$\times$10$^{-10}$	\\
Y3   & 122 & $2564\pm53$  & 3.6 &  4.6$\times$10$^{-9}$   \\
Y2   & 122 & $2330\pm50$  & 5.1 &  6.4$\times$10$^{-9}$\\
Y1   & 122 & $2093\pm42$  & 2.4 &  3.0$\times$10$^{-9}$\\

\hline

\end{tabular}

\end{minipage}
~ \footnotesize{$^a$ Tangential velocities from \cite{jochen94_hh46}. $^b$ \h\ temperature. $^c$ \h\ column density. $^d$ \h\ mass ejection rate}
\end{table}


\subsection{H$_2$ temperature and mass flux}

\begin{figure}[!ht]
\begin{center}
\includegraphics[totalheight=8cm]{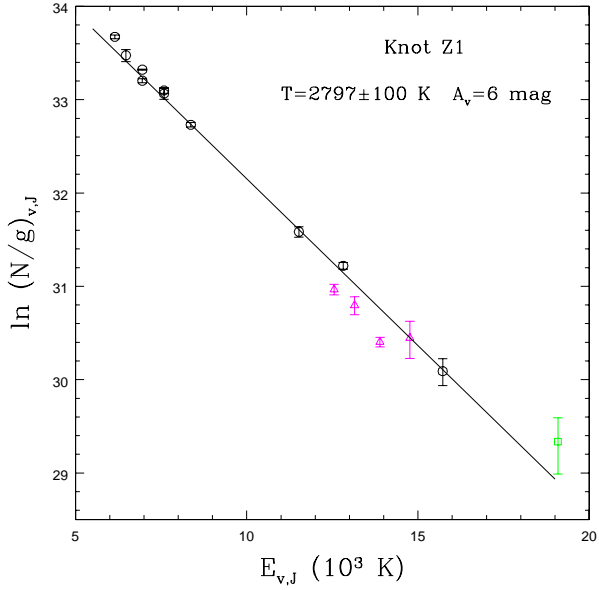}
\caption{\label{fig:rot_z2}
H$_2$ rotational diagram for the different lines in the knot Z1 of HH46-47. Different symbols indicate lines coming from different vibrational levels: circles indicate v=1, triangles v=2 and squares v=3. The straight line represents the best fit through the data. The corresponding temperature and the adopted extinction value is indicated.}
\end{center}
\end{figure}

Using the different \h\ detected transitions, excitation diagrams for several knots (Z1, X, Y3, Y2 and Y1) along the jet showing enough \h\ lines have been constructed.
As an example, Fig.\,\ref{fig:rot_z2} shows the excitation diagram of the knot Z1, the brightest knot observed in HH46-47. The excitation diagrams for the rest of the knots are reported in the Appendix\,\ref{app:rovibrational}. In this kind of diagrams the natural logarithm of $N(\nu\,,J)/g_{\nu\,,J}$ versus the excitation energy $E(\nu\,,J)$ of a given level $(\nu\,,J)$ is represented, where $N(\nu\,,J)/g_{\nu\,,J}$ is the ratio of the column density over the statistical weight.
The different symbols indicate lines coming from different vibrational levels, while the straight line represents the best linear fit through our data. The slope of the line is proportional to the inverse of the gas temperature.
Extinction values as derived from the \fe\ lines have been adopted. The extinction can, in principle, be also derived from the \h\ 1-0S(1)/1-0Q(3) line ratio. However, this method involves lines close in wavelength and thus it is not sensitive enough to extinction variations. In addition, the Q lines fall in a poor atmospheric transition region (around $\sim$2.3\,\um) being their fluxes affected by large uncertainties. The fact that our data can be well traced by a straight line on the rovibrational diagram indicates, however, that the extinction values derived from the \fe\ lines are in good agreement with the \h\ ones.
The resulting temperature values are indicated in Table 4 and range between $\sim$2000\,K and $\sim$2800\,K.

The mass flux carried out by the warm molecular component (\mh) has been also estimated from the observed \h\ transitions in the knots Z1, X, Y3, Y2 and Y1. \mh\ can be inferred from the relationship \mh=$2\, N(H_2)\, A\, dv_t / dl_t$. Here, A is the area of the emitting region sampled by the slit, $dl_t$ and $dv_t$ are the projected length and the tangential velocity, and $N(H_2)$ is the total column density.
\mh\ has been derived in the same spatial region extracted for the \fe\ line analysis. The $dv_t$ value has been taken from \cite{jochen94_hh46}, while $N(H_2)$ for each knot has been found from the intercept to zero of the straight line fitted to the \h\, transitions in the Boltzmann diagram.
The derived values for each knot are reported in Table\,\ref{tab:h2}. The average \mh\ value along the jet is $\sim3.8\times10^{-9}$\,M$_{\odot}$\,yr$^{-1}$. 
This estimate is however a lower limit since we are only considering the warm component of the gas.

\subsection{Electron density}

The electron density (\Ne ) can be derived from \fe\, lines coming from levels with similar excitation energies. This is the case of the ratio \fe\,1.600/1.644\,\um\,, which is sensitive to \Ne\ values between $10^3-10^5 cm^{-3}$ and weakly dependent on the temperature . This ratio has been fitted with a NLTE code including the first 16 fine-structure levels of the \fe\ (\citealt{nisini02}). 
\begin{table}
\begin{minipage}[t]{\columnwidth}
\caption{Electron density along the HH46-47 jet.}
\label{tab:ne}
\centering
\renewcommand{\footnoterule}{}  
    \begin{tabular}[h]{cc|c|cc}
 \hline     \hline \\[-5pt]

Knot 
& r$_t$\footnote{Distance from the source in arcsec. Negative values correspond to the southeastern, redshifted jet axis.} &
n$_e\pm \bigtriangleup$n$_e$\footnote{Electron density measured from the ratio \fe\,1.60/1.64\,\um\,.}  &
n$_e$(HVC)\footnote{Electron density measured from the ratio \fe\,1.60/1.64\,\um\, for the HVC and LVC.} &
n$_e$(LVC)$^b$\\

 & (\arcsec\,) & (cm$^{-3}$) & (cm$^{-3}$)   &  (cm$^{-3}$) \\
\hline \\[-5pt]
S    & (-3.4,+2.8)  & 6300$^{+1700}_{-1300}$ &	5400	&	7650	 \\[+5pt]
Z2   & (-9.3,-5)&  1300$^{+1800}_{-1200}$   &	2350	&	4600	\\[+5pt]
Z1   & (-14.8,-9.3)   & 3200$^{+700}_{-850}$  &	3100	&		\\[+5pt]
\hline
\end{tabular}
\end{minipage}
\end{table}

The electron density has been derived integrating the total flux of the line profile of every extracted knot for both lines. The \Ne\, values and the intervals used to extract the knots are reported in Table\,\ref{tab:ne}. 
Previous studies on protostellar jets, like HH1, HH111 and HH34, have shown evidences of a decrease of the electron density with distance (see, e.g., \citealt{BE,nisini_hh1,linda06}). The same behaviour may also occur in HH46-47. Indeed the electron density roughly decreases from a value of 6300\,cm$^{-3}$ in knot S to a value of 3200\,cm$^{-3}$ in knot Z1. These two knots are displaced, however, in different lobes of the jet. In addition, the increase of \Ne\ from knot Z2 to knot Z1 and the the lack of further \fe\ emitting knots prevent us to properly probe the dependence on distance of \Ne. 



\begin{figure}[!ht]
\begin{center}
\includegraphics[totalheight=6cm]{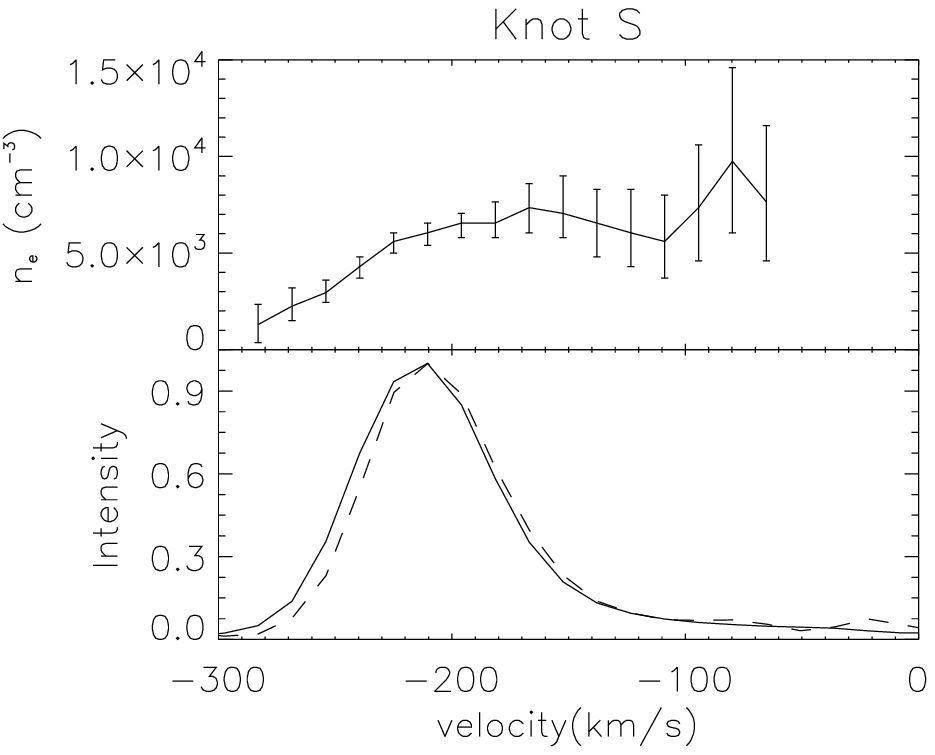}
\includegraphics[totalheight=6cm]{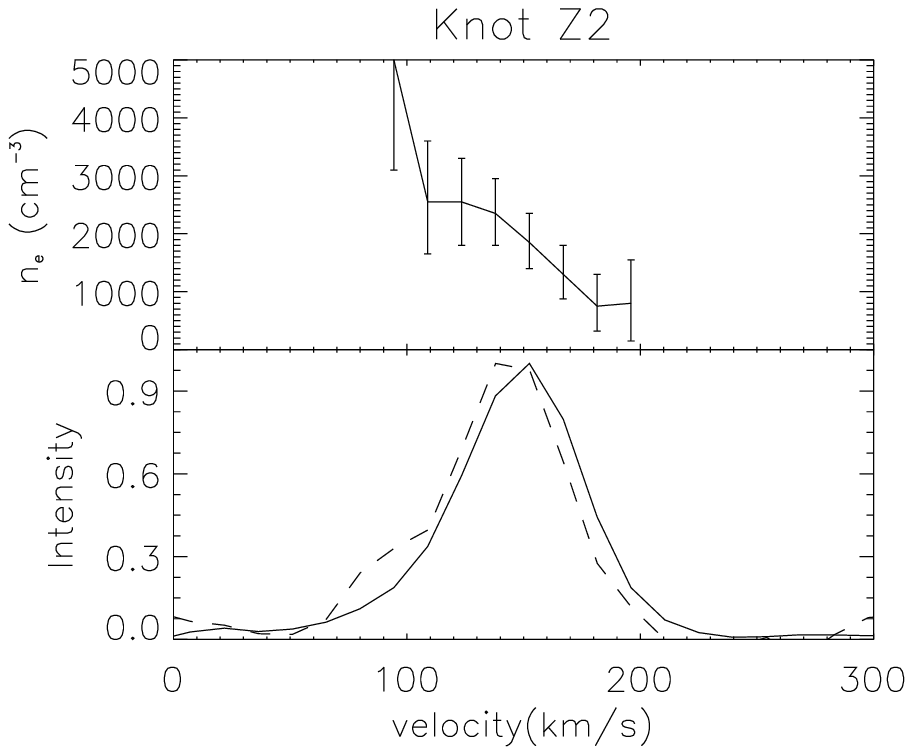}
\includegraphics[totalheight=6cm]{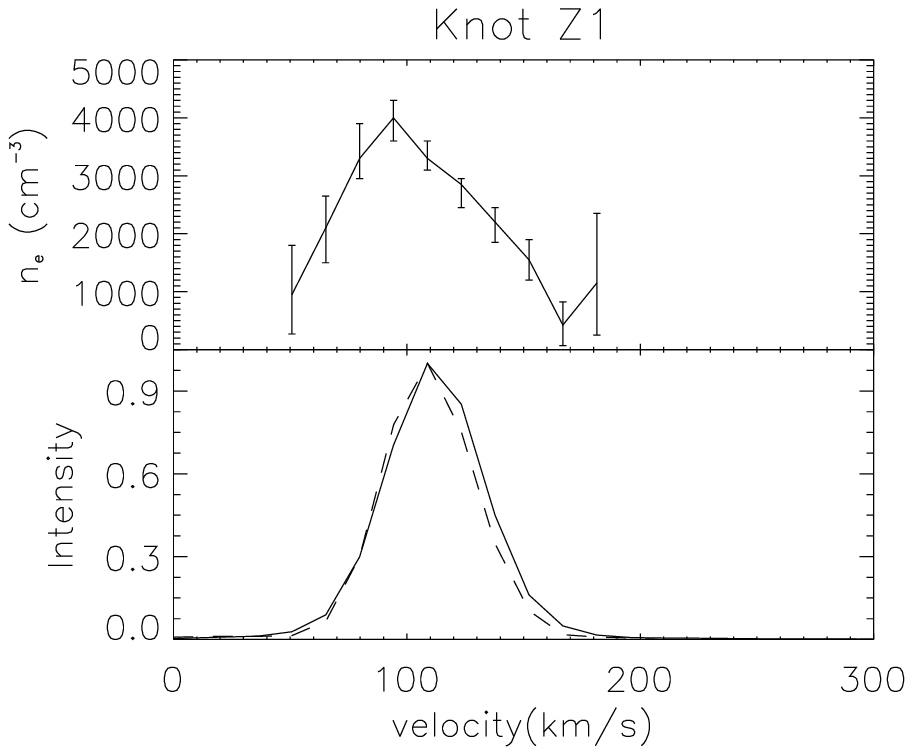}
\caption{\label{fig:ne}
Normalised to one line profiles (lower panels) of the \fe\, lines 1.644\,\um\, (solid line) and 1.600\,\um\, (dotted line), and the electron density in each velocity channel (upper panels) for different extracted knots along the HH46-47 jet.}
\end{center}
\end{figure}

\subsection{\fe\ depletion}

Refractory such species as iron are expected to be locked into dust grains in the quiescent conditions of molecular clouds. High velocity shocks inside the jet beam are, however, supposed to destroy the grains and  release Fe atoms in gaseous form. In recent years several studies have been done in order to quantify the presence of dust along protostellar jets and measure the shock efficiency in disrupting the dust grain cores (see  e.g., \citealt{nisini_hh1, linda06,teresa08}).
The iron gas phase abundance can be measured comparing the emission of \fe\ with that of a non-refractory species of known abundance. A very suitable ratio is the one between the \fe\,1.257\,\um\ line and the \pii\,1.188\,\um\ line \citep{oliva01}. These two lines are very close in wavelength and thus their ratio is poorly affected by extinction. In addition, both lines are excited under similar conditions and can be assumed in the first ionised state at similar percentage. Assuming a solar abundance for both elements, the expected theoretical line ratio is \fe\,/\,\pii$\sim$1/2\,[Fe\,/\,P]$\sim$56 \citep{oliva01}. 
Thus, observed values lower than this  indicate that some of the iron is locked in grains.

Along the HH46-47 jet, both the \fe\,1.257\,\um\, and the \pii\,1.188\,\um\ lines have been detected in knots S and Z1. The measured \fe\ / \pii\ ratio is $\sim$7.3 in knot S and $\sim$25.6 in knot Z1. These values indicate that around 88\% and 58\% of iron is still  in dust grains in knots S and Z1, respectively. This leads to a ratio (Fe/H)$_{gas}\sim3.92\times10^{-6}$ in the knot S and a (Fe/H)$_{gas}\sim1.18\times10^{-5}$ in the knot Z1. The strong iron depletion at the jet base is still lower than the 99\% diffuse gas iron depletion (see, e.g., \citealt{savage96}), indicating that a small fraction of the iron have come back into gas phase at the base of the jet with respect to the interstellar value.
Evidence that the gas phase iron abundance is low in knots close to the source and increases with
the distance have been also found in the HH1 jet (\citealt{nisini_hh1}). 

In knot Z1, the \Ti\,1.140\,\um\ line has been also detected. Ti is expected to be fully ionised in the jet plasma, since it has a low ionisation potential of 6.82\,eV. As iron, the titanium is a high-refactory element and thus the Fe/Ti ratio can give us some clues about the relative abundance of both elements, as well as to the selective depletion pattern of both elements in the dust along protostellar jets. With this aim we have computed the \fe\,1.257\,\um / \Ti\,1.140\,\um\ intensity ratio using a NLTE model for these species \citep{rebeca08}. 
The observed ratio, assuming the physical conditions present in knot Z1 is $\sim$70, indicating a Fe/Ti ratio 5 times smaller than the theorically expected of [Fe/Ti]$_\odot$ = 354 using the solar abundances of \cite{asplund05}. 
The small Fe/Ti is consistent with what found in the HH1 jet \citep{rebeca08}, where a Fe/Ti ratio 2-3 times lower than solar was derived.

\section{Medium resolution spectral analysis}

\subsection{Electron density and Mass flux}

The \fe\,1.600 and 1.644\,\um\, lines have been also detected at medium resolution in the knots S, Z1 and Z2. In these knots, we can test 
how  the electron density varies as a function of the radial velocity, by measuring  the 1.600/1.644\,\um\, ratio in each pixel along the spectral profile. 
The \fe\, ratio was computed only for the velocity points where the intensity is greater than S/N $\sim$ 3. 
Figure\,\ref{fig:ne} shows electron density variations along the MR spectral profile of the knots S, Z1 and Z2 (upper panels) together with the normalised line profile of the \fe\,1.600\,\um\ and 1.644\,\um\, lines (lower panels).  The electron density increases as the velocity decreases in all of the knots where two velocity components were detected in \fe. In particular, the knot S close to the source has a behaviour very similar to what was found in HH34 (\citealt{rebeca08}), with the electron densities higher in the gas at low velocity.
We have separately measured the \Ne\ in the HVC and LVC components. Near to the source (knot S),  we have defined a HVC ranging from $\sim$-239\,\kms\, to $\sim$-181\,\kms\  (consistent with the FWHM of a Gaussian-fit
through the line), and a LVC from $\sim$-181\,\kms\, to $\sim$-50\,\kms. Far from the source, only the knot Z2 shows a clear double
velocity component: here we have defined the intervals from $\sim$110\,\kms\, to $\sim$167\,\kms\, and from $\sim$80\,\kms\, to $\sim$110\,\kms\, for the high and low velocity gas, respectively.

The \Ne\ values for the integrated velocity components are reported in Table\,\ref{tab:ne}. 
The electron density of the LVC (\Ne(LVC)$\sim$4600-7600) is larger than the electron density of the HVC (\Ne(HVC)$\sim$2300-5000) as evidenced by Fig.\,\ref{fig:ne}. 
The  behaviour found in knot S is similar to what we have already found at the HH34 jet base, and contrast with 
the results on T Tauri stars, where the highest electron density is associated with the high velocity gas pertaining
to the more collimated inner jet section, while the low velocity component is a less collimated gas at lower density (e.g., \citealt{coffey08}). These contrasting results may reflect a different excitation conditions in the structure of Class I jets, or may be a result biased by the low 
spatial resolution of our observations that does not allow us to resolve the same spatial scales probed in T Tauri jets, namely within 100-200\,AU.

At the bow shock positions far from the source, like in the Z2 knot, variations of the electron density with velocity may depends on the
bow inclination angle and section of the bow intercepted by the slit, as shown by \cite{indebetouw95}. According to this model, a bow shock seen on-axis at an inclination angle of $\sim$30 degree, as in our case, shows qualitatively a behaviour like the one we
observe, namely with the highest electron densities displaced at low velocity.

\begin{table}
\begin{minipage}[t]{\columnwidth}
\caption{\mjet\ along the HH46-47 jet.}
\vspace{0.2cm}
\label{tab:mflux}
\centering
\renewcommand{\footnoterule}{}  
    \begin{tabular}[h]{c|cc|cc}
 \hline     \hline \\[-5pt]
Knot 
& dv$_t$(HVC)\footnote{Tangential velocity derived from the observed radial velocities assuming an orientation angle of the jet with respect to the plane of the sky of 34\degr\ (\citealt{jochen94_hh46}).}
& \mjet(HVC)\footnote{\mjet\ measured from the luminosity of the \fe\,\,1.64\,\um\ for the HVC and LVC. The lower value has been derived assuming an iron solar abundance, while the upper value has been inferred assuming the iron gas-phase abundance derived from the \fe/\pii\ ratio.} 
& dv$_t$(LVC)$^a$
& \mjet(LVC)$^{b}$  \\  
  &  \small{\kms}  & \small{10$^{-7}$\,\msyr} & \small{\kms} & \small{10$^{-8}$\,\msyr} \\ 

\hline \\[-5pt]
S &  316 & 0.3-1.9 & 205 & 0.5-3.6 \\
Z2 & 213 & 0.4-0.9 & 136 & 0.2-0.4  \\
Z1 &  166 & 0.4-1.0 &     &  		\\

\hline
\end{tabular}
\end{minipage}
\end{table}


In order to compute the contribution to the mass flux of the atomic component, we have derived \mdot\, from the luminosity of the \fe\,1.644\,\um\, line, following the approach described in \cite{rebeca08}. To compute the theoretical \fe\, lines fractional population, the \Ne\ values derived separately
for the LV and HV components have been assumed. A constant value of 10\,000\,K have been instead assumed for the electron temperature:
the derived populations are only a little sensitive to this value and the results do not change significantly if the temperature is maintained
in between $\sim$ 7000\,K and 15\,000\,K.
An inclination angle of the jet with respect to the plane of the sky, $i\sim34\degr$ (\citealt{jochen94_hh46}), has been assumed and used to infer the velocity of the knots projected perpendicular to the line of sight through our measurements of the radial velocity. 
These values are reported in Table\,\ref{tab:mflux}, while the adopted length of the knots projected perpendicular to the line of sight is 
listed in Table\,\ref{tab:ne}. The \fe\,1.644\,\um\ line flux has been dereddened using the extinction derived from the ratio \fe\,1.644/1.257 and reported in Table\,\ref{tab:Av}. 

\mjet\ values have been derived both assuming a solar abundance of iron and considering the gas-phase Fe abundance measured from the Fe/P analysis, in the assumption that the same dust depletion factor is valid for both the HVC and LVC. The corresponding range of values 
are listed in Table\,\ref{tab:mflux}.
Values of the order of 0.5-1$\times$10$^{-7}$\,\msyr\ are found in the HVC, while the LVC shows $\sim16$ times lower values.


\begin{figure}
  \resizebox{\hsize}{!}{\includegraphics{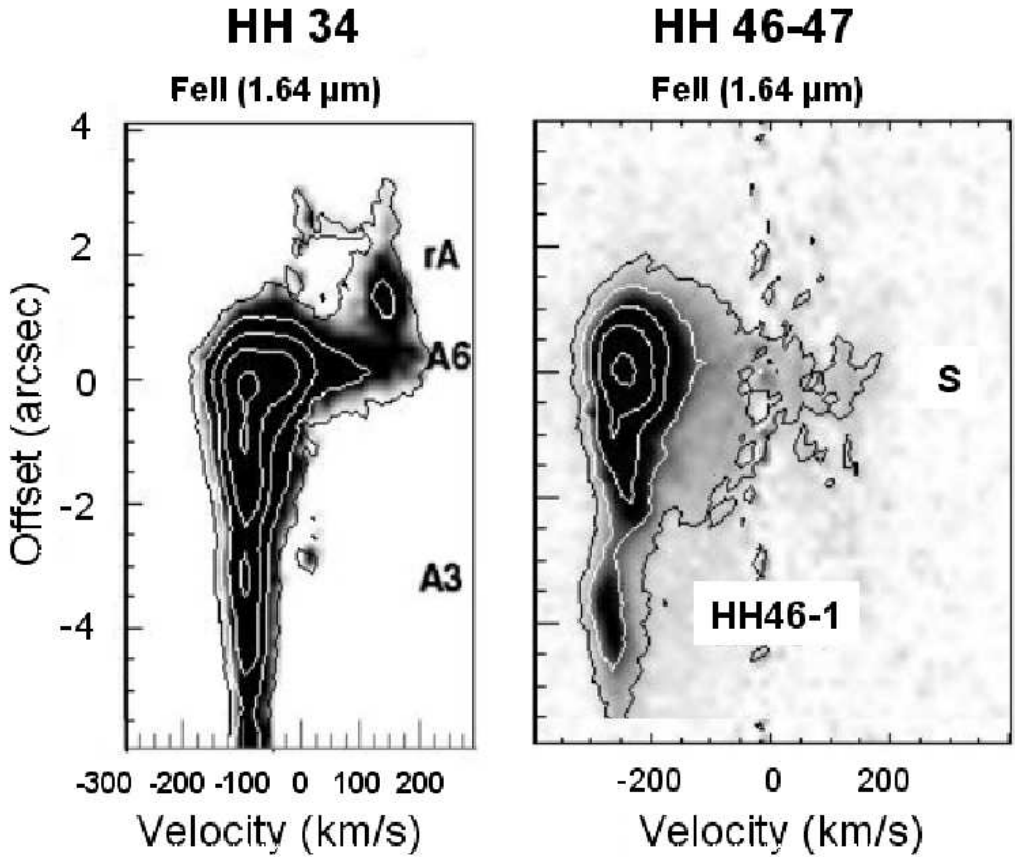}}
  \caption{PVD of the \fe\,1.644\,\um\ line for the HH34 and HH46-47 jets.}
  \label{fig:comparison}
\end{figure}

\section{Discussion}

The \fe\ and \h\ emission structure at the base of HH46-47 is very similar to the one observed in several Class I jets (e.g., \citealt{davis_MHEL,davis03,takami}). As in the case of CTTS jets, the Forbidden-Emission Line (FEL) regions of Class I jets are characterised by the presence of two velocity components at the jet base at high and low velocity. The MHEL regions, as traced by \h\ emission, show on contrast just a LVC near the source while higher velocity \h\ is usually detected further downsteam.
In particular, the kinematics and physical parameters of the HH46-47 jet are very similar to those found in our previous work about HH34 \citep{rebeca08} where the physical properties have been also derived as a function of velocity.
For instance, Fig. \,\ref{fig:comparison} shows the \fe\,1.644\,\um\ PVDs of the inner region of these two jets. Since the two objects are at a similar distance ( i.e. 450 pc), the emission regions have the same physical extent in this figure. 
In both the PVDs, the \fe\ lines broaden and emission at lower velocities, down to 0\,\kms\ appears within $\sim$2\arcsec\ from the central source, i.e. up to distances of $\sim$1000\,AU. The extend of the LVC is comparable with the typical size of the envelope of Class I sources (see, e.g., \citealt{andre94}). Thus this result may indicate an interaction of a low-velocity disc wind with the source envelope as the possible origin of the LVC in Class I jets. The large extend of the LVC is at variance with the PVDs of most of the CTT, where the LVC is usually confined within $\sim$ 200\,AU from the source \citep{pyo03,pyo06}.

Also the values of electron density and mass flux rate that we derive in HH46-47, as well as their dependence with velocity are similar to those inferred on HH34, i.e., $n_e$ $\sim$ 0.5-1$\times$10$^4$ cm$^{-3}$ in the HVC and 
$\sim$1-2$\times$10$^4$ cm$^{-3}$ in the LVC, and \mjet\, of the order 10$^{-7}$ \msyr. Similar values 
have been also estimated on a number of Class I YSO jets (e.g., \citealt{davis03,linda06,takami}).

At variance with \fe, the H$_2$ PVDs of HH46-47 shows a different behaviour with respect to the one of HH34. 
In HH34 the \h\ PVD shows spatially- and kinematically-separated LVC and HVC, and only the LVC is visible down to the central source. 
In HH46-47,on the contrary, a single velocity component near the source is detected. The morphology and kinematics of the \h\  close to the source (the so-called Molecular Emission Line, MHEL, regions; \citealt{davis_MHEL})  in class I objects varies indeed 
significantly from source to source \citep{davis_MHEL,davis02} and this is another element that makes the origin of the MHEL regions unclear. 

In \cite{rebeca08} we have compared the \fe\,1.644\,\um\ PVDs of HH34 with those of the synthetic PVDs available for the disc-wind and X-wind models \citep{pesenti04,shang98}. Most of this  discussion is also valid for the HH46-47 jet. 
Briefly, none of these models are able to explain the persistency of the LVC at large distances from the launching regions. 
This may suggest that the low velocity that we observe far from the source represent dense gas entrained by the high-velocity 
collimated jet, rather than gas directly ejected from the acceleration region.

From the observed HH46-47 peak radial velocity of the HVC, corrected for the inclination angle, we estimate a jet 
poloidal velocity around 350\,\kms.  This value is just at the border of the range of poloidal velocities predicted by X-wind
models and it is also  in agreement with disc-wind models with  launching radii in the range 0.07-0.15 AU and magnetic lever arms
larger than $\sim$ 5 \citep{ferreira06}. 

Further constraints on MHD launching models may be derived from the presence of  dust in the initial beam of the  HH46-47 jet, as suggested 
by \cite{linda09}.
We have in fact seen that around 88\% of the iron within 450\,AU from the source is  depleted, implying 
that a large fraction of dust grains are still present at the jet base. Dust grains of the ambient medium are expected to be destroyed 
by the passage of the initial  leading bow-shock, that have  velocities higher than 100\,\kms\ \citep{draine03}. 
Since the amount of iron depletion decreases toward the external knots (only 58\% of iron is found to be depleted in grains at $\sim$5000\,AU from the source) dust grains are unlikely to come from entrained material. In addition they cannot be reformed along the jet beam since the time required for grain reformation is $\sim$10$^7$-10$^8$\,yr \citep{jones01}, while the jet takes $\sim$1300\,yr to travel from the central source to the final bow-shock \citep{hartigan05}. Therefore, the larger amount of grains near the source may suggest that they have been removed from the accretion disc and transported through the jet.
Interestingly it has been showed that, in the case of Herbig stars and TTauri stars, the disc is populated by dust grains only beyond the so-called dust evaporation point, R$_{evp}$ \citep{akeson05,eisner_tt07, eisner_herbig07}. Future similar measurements of R$_{evp}$ for Class 0/I sources would help to constraint the region from which dusty jets are launched from.


\section{Conclusions}

We have presented medium resolution (MR) (in the H- and K- band) and low resolution (LR)  (in the  J-, H- and K-band) spectra of the
Class I driven  HH46-47 jet. These observations allow us to resolve two velocity components at the jet base and to study the kinematics and physical properties of the jet as a function of velocity. From the \fe\,1.644/1.600\,\um\ lines the electron density has been derived along the jet and as a function of velocity. In addition, from the luminosity of the \fe\,1.644\,\um\ line the mass flux for the different velocity components has been derived. Finally, the \fe\ / \pii\ ratio has been used to retrieve important information about the dust depletion inside the jet beam. We summarise our results as follows:
\begin{itemize}
\item We were able to trace the HH46-47 jet down to the central source. While the atomic component shows a HVC and LVC near the source, only a single velocity component has been detected for the molecular gas. Within 1\arcsec\ from the source, the atomic emission broaden down to 0\,\kms\ and even reaches red-shifted velocities.

\item From the ratio of the \fe\,1.644, 1.257\,\um\ lines detected in the LR spectra, the visual extinction for different knots have been derived. We have found an A$_V$ ranging from $\sim$6\,mag to $\sim$12\,mag along the jet.

\item Several H$_2$ lines have been detected from the LR spectra allowing us to construct excitation diagrams for different knots. From these diagrams we have derived the temperature and mass flux for the warm molecular component of the jet. The temperature has an average value around $\sim$2300\,K, while we have found an average mass flux for the warm molecular component of $\gtrsim 3.7\times10^{-9}$\,M$\odot$\,yr$^{-1}$.

\item The electron density, derived from the  \fe\,1.600/1.644\,\um\ ratio, roughly decreases from the internal to the more external knots from 9400 to 3200\,cm$^{-3}$.  Larger electron density values are found in the LVC with respect to the HVC for all the extracted knots, with an average value of $\sim$6000 and $\sim$4000 cm$^{-3}$ for the LVC and HVC, respectively.

\item  From the \fe\ / \pii\ ratio, we have estimated the iron depletion in the knots S and Z1. We have found that around 88\% and 58\% of the iron is depleted in grains in the knot S and Z1. 

\item The mass flux rate transported by the \fe\ has been derived for both the velocity components assuming the Fe gas-phase abundance derived in this paper. We found that the HVC is transporting the bulk of the mass along the jet, with average values of \mjet(HVC)$\sim$1.2$\times$10$^{-7}$\,\msyr\ and \mjet(LVC)$\sim$2$\times$10$^{-8}$\,\msyr. As in the case of HH34 (\citealt{rebeca08}), the difference in tangential velocity between the HVC and LVC is not enough to account for the much hihger \mjet\ value in the HVC, indicating a significantly larger value of the (n$_H$V)\footnote{where n$_H$ and V are the total density and the emitting volume, see \cite{rebeca08} for more details.} product in the HVC than in the LVC (from a factor $\sim$1.5 in the knot Z2 to $\sim$3.5 in knot S).

\item Many of the derived properties of the HH46-47 are common to jets from YSOs in different evolutionary states. The derived densities and mass flux values are typical of Class I objects or very active T Tauri stars. However, the observed spatial extent of the LVC and the velocity dependence of the electron density are not explained by the current jet formation models.

\end{itemize}

\begin{acknowledgements}
The present work was partly supported by the European Community’s Marie Curie Actions - Human Resource and Mobility within the JETSET (Jet Simulations, Experiments and Theory) network under contract MRTN-CT-2004 005592. 
L.P. work was funded by a Fellowship from the Irish Research Council for Science, Engineering and Technology.
\end{acknowledgements}

\bibliographystyle{aa}
\bibliography{references}

\Online

\begin{appendix}
\onecolumn
\section{knot fluxes at low resolution}
\label{app:fluxes}

\begin{table}[!h]
\begin{minipage}[t]{\textwidth}
\caption{Observed unreddened lines along HH46-47: atomic lines.}
\label{tab:atomic_fluxes} 
\centering
\vspace{0.1cm}
\renewcommand{\footnoterule}{}  
    \begin{tabular}[h]{ccccccc}
 \hline     \hline \\[-5pt]

Line id. & $\lambda$(\um) & \multicolumn{5}{c}{F ($\Delta$ F)(10$^{-16}$erg s$^{-1}$cm$^{-2}$)} \\
     	&		  	           &   S      &	HH46-1	 & B8	    & Z2       & X	  \\
\hline \\[-5pt]
\pii\,$^3$P$_{2}$-$^1$D$_{2}$      & 1.188 & 6.2(1.2) & ...      & ...      & ...      & ...      \\
\fe\,a$^4$D$_{5/2}$-a$^6$D$_{7/2}$ & 1.249 & 1.9(0.6) & ...      & ...      & ...      & ...      \\
\fe\,a$^4$D$_{7/2}$-a$^6$D$_{9/2}$ & 1.257 & 54.2(0.8)& 15.8(0.9)& 11.9(0.7)& 7.1(0.9) & 6.4(1.0) \\
\fe\,a$^2$G$_{9/2}$-a$^4$D$_{7/2}$ & 1.267 & 3.2(0.8) & ...      & ...      & ...      & ...      \\
\fe\,a$^4$D$_{1/2}$-a$^6$D$_{1/2}$ & 1.271 & 2.0(0.3) & ...      & ...      & ...      & ...      \\
\fe\,a$^4$D$_{3/2}$-a$^6$D$_{3/2}$ & 1.279 & 2.4(0.6) & ...      & ...      & ...      & ...      \\
\fe\,a$^4$D$_{5/2}$-a$^6$D$_{5/2}$ & 1.295 & 8.9(1.1) & ...      & ...      & ...      & ...      \\
\fe\,a$^4$D$_{3/2}$-a$^6$D$_{1/2}$ & 1.298 & 3.4(0.9) & ...      & ...      & ...      & ...      \\
\fe\,a$^2$G$_{7/2}$-a$^4$D$_{3/2}$ & 1.300 & 1.7(0.6) & ...      & ...      & ...      & ...      \\
\fe\,a$^4$D$_{7/2}$-a$^6$D$_{7/2}$ & 1.321 & 13.1(0.4)& 4.5(1.4) & 3.5(0.5) & 3.0(1.2) & 2.4(1.2) \\
\fe\,a$^4$D$_{5/2}$-a$^6$D$_{3/2}$ & 1.328 & 4.3(0.6) & ...      & ...      & ...      & ...      \\
\fe\,a$^4$D$_{5/2}$-a$^4$F$_{9/2}$ & 1.534 & 9.9(1.0) & ...      & ...      & 0.7(0.4) & ...      \\
\fe\,a$^4$D$_{3/2}$-a$^4$F$_{7/2}$ & 1.600 & 11.2(1.0)& ...      & ...      & 0.9(0.4) & ...      \\
\fe\,a$^4$D$_{7/2}$-a$^4$F$_{9/2}$ & 1.644 & 85.8(0.7)& 8.9(0.5) & 8.7(0.8) & 20.1(0.5)& 4.4(0.5) \\
\fe\,a$^4$D$_{1/2}$-a$^4$F$_{5/2}$ & 1.664 & 5.7(0.8) & ...      & ...      & 1.1(0.7) & ...      \\
\fe\,a$^4$D$_{5/2}$-a$^4$F$_{7/2}$ & 1.677 & 13.3(0.9)& ...      & ...      & 1.8(0.4) & ...      \\
\fe\,a$^4$D$_{3/2}$-a$^4$F$_{3/2}$ & 1.797 & 5.2(0.9) & ...      & ...      & ...      & ...      \\
\fe\,a$^4$D$_{5/2}$-a$^4$F$_{5/2}$ & 1.800 & 14.3(0.7)& ...      & ...      & ...      & ...      \\
\fe\,a$^4$D$_{7/2}$-a$^4$F$_{7/2}$ & 1.810 & 52.6(1.1)& ...      & ...      & ...      & ...      \\
\fe\,a$^2$P$_{3/2}$-a$^4$P$_{3/2}$ & 2.133 & 3.2(0.5) & ...      & ...      & ...      & ...      \\
\hline
    \end{tabular}
\end{minipage}

\end{table}

\begin{table}
\begin{minipage}[t]{\textwidth}
\caption{ Observed unreddened lines along HH46-47: H$_2$ lines.}
\label{tab:molecular_fluxes}
\centering
\vspace{0.1cm}
\renewcommand{\footnoterule}{}  
\begin{tabular}[h]{cccccccc}
 \hline     \hline \\[-5pt]

Line id. & $\lambda$(\um) & \multicolumn{6}{c}{F ($\Delta$ F)(10$^{-16}$erg s$^{-1}$cm$^{-2}$)} \\
     	&		&   S      & Z2       & X	 & Y3	    & Y2        &Y1	  \\
\hline \\[-5pt]
H$_2$ 2-1 S(5) & 1.945  & ...	   & ...      & ...      & ...      & ...      & 2.7(0.7)	\\
H$_2$ 1-0 S(2) & 2.034 & ...	   & 2.1(0.6) &	26.6(0.6)& 10.0(0.6)& 13.2(0.4)& 8.7(0.4)		  \\
H$_2$ 2-1 S(3) & 2.074 & ...	   & ...      &	3.7(0.5) & 2.3(0.2) & 3.9(0.4) & ...			\\
H$_2$ 1-0 S(1) & 2.122 & 12.0(0.5) & 5.6(0.3) &	36.3(0.3)& 27.0(0.3)& 3.3(0.2) & 22.7(0.2)	\\
H$_2$ 2-1 S(2) & 2.154 & ...	   & ...      &	1.1(0.2) & ...      & 1.6(0.2) &...		\\
H$_2$ 1-0 S(0)\footnote{Blend with \fe\ b$^2$F$_{5/2}$-a$^2$F$_{7/2}$} 
	       & 2.224 & 12.7(0.6) & 3.1(0.6) &	7.7(0.4) & 5.4(0.5) & 6.9(0.3) & 5.2(0.2)  \\
H$_2$ 2-1 S(1) & 2.248 & ...	   & 1.4(0.4) & 3.4(0.3) & 3.0(0.5) & 4.6(0.2) & 3.5(0.3) \\
H$_2$ 1-0 Q(1) & 2.407 & 16.6(1.3) & 4.8(0.6) &	28.9(0.6)& 20.8(0.8)& 24.3(0.7)& 17.8(0.5)	\\
H$_2$ 1-0 Q(2) & 2.413 & 13.8(1.5) & 2.3(0.3) & ...      & ...      & ...      & ...	\\
H$_2$ 1-0 Q(3) & 2.424 & 16.2(1.3) & 5.8(0.5) &	30.7(0.6)& 23.3(1.5)& 26.5(0.7)& 17.8(0.5) 	\\
H$_2$ 1-0 Q(4) & 2.437 & 9.3(1.1)  & ...      &	8.4(0.7) & 6.1(1.2) & 8.0(0.6) & 5.2(0.6)	\\
H$_2$ 1-0 Q(5) & 2.455 & ...       & ...      &	23.4(0.9)& 15.7(1.6)& 21.6(0.8)& 14.9(0.7) \\
H$_2$ 1-0 Q(6) & 2.476 & ...       & ...      &		 & ...      & 6.6(0.6) & ...	\\
\hline

\end{tabular}

\end{minipage}

\end{table}



\newpage
\end{appendix}

\begin{appendix}

\section{Rovibrational diagrams}
\label{app:rovibrational}

\begin{figure}[!h]

\includegraphics[totalheight=7.2cm]{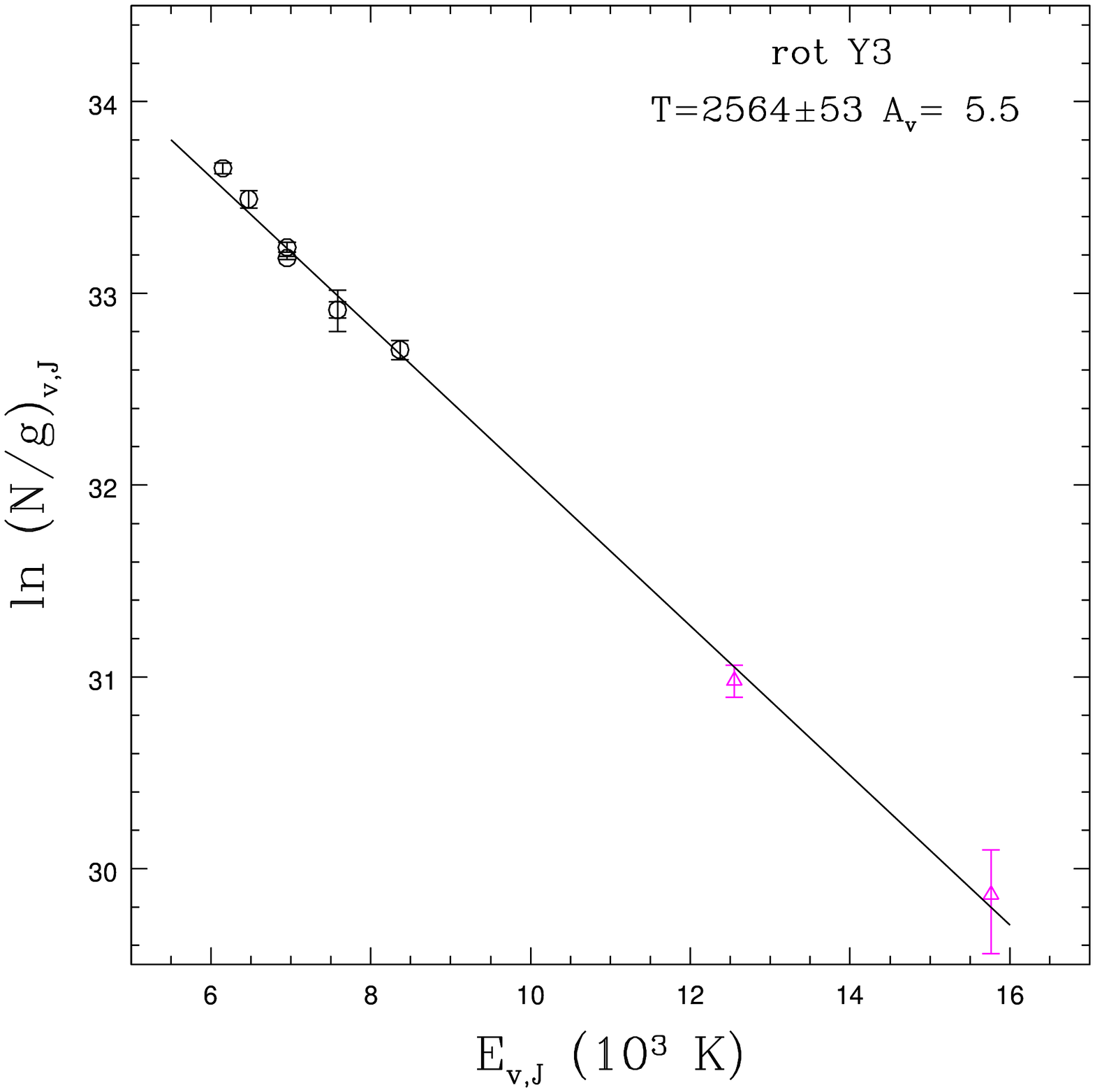}
\centering
\caption{\label{knot_y3} As Fig.\,\ref{fig:rot_z2} for knot Y3.}
\end{figure}

\begin{figure}[!h]
 
\includegraphics[totalheight=7.2cm]{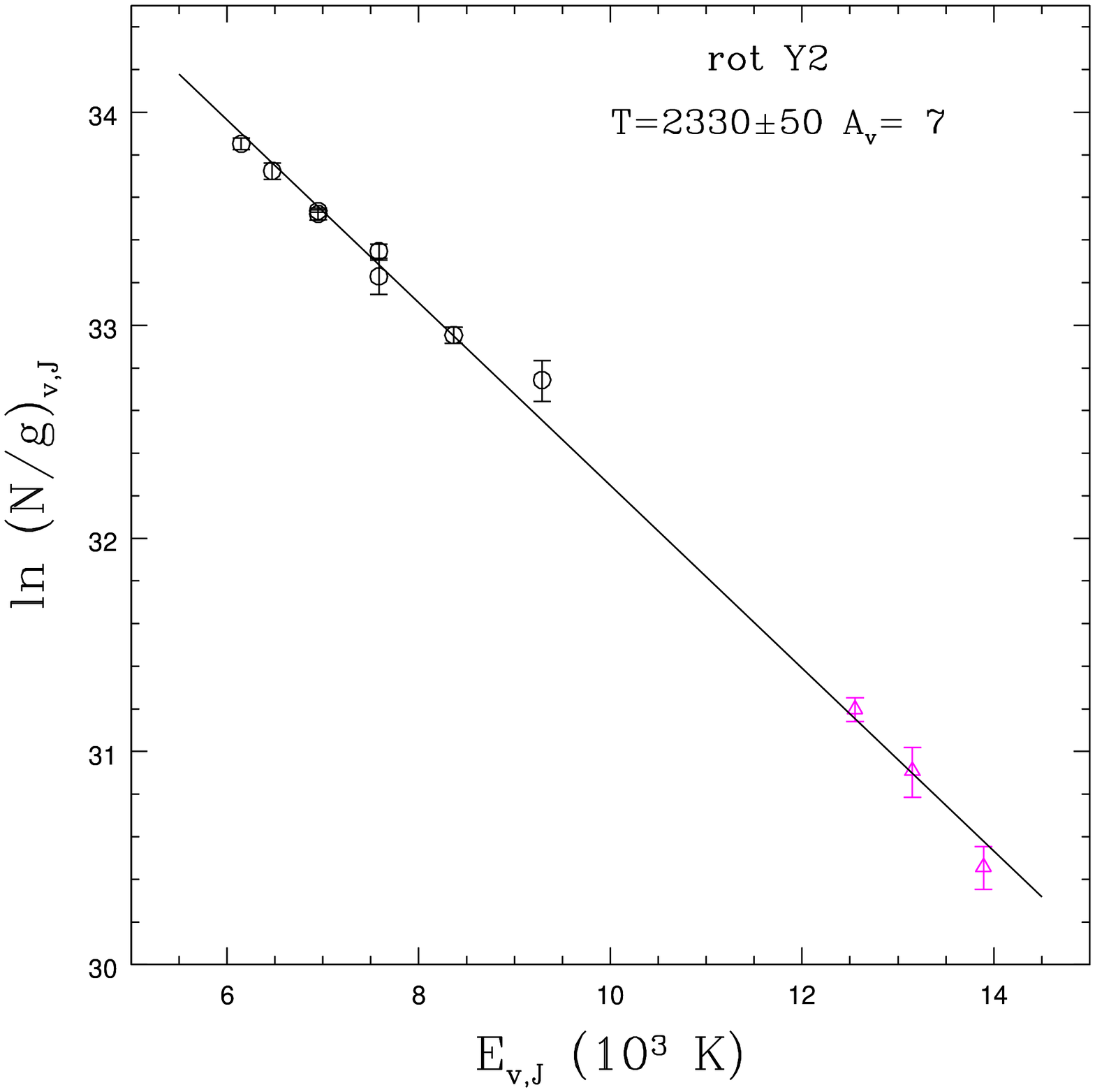}
\centering
\caption{\label{knot_y2}As Fig.\,\ref{fig:rot_z2} for knot Y2}
\end{figure}

\begin{figure}[!h]
 
\includegraphics[totalheight=7.2cm]{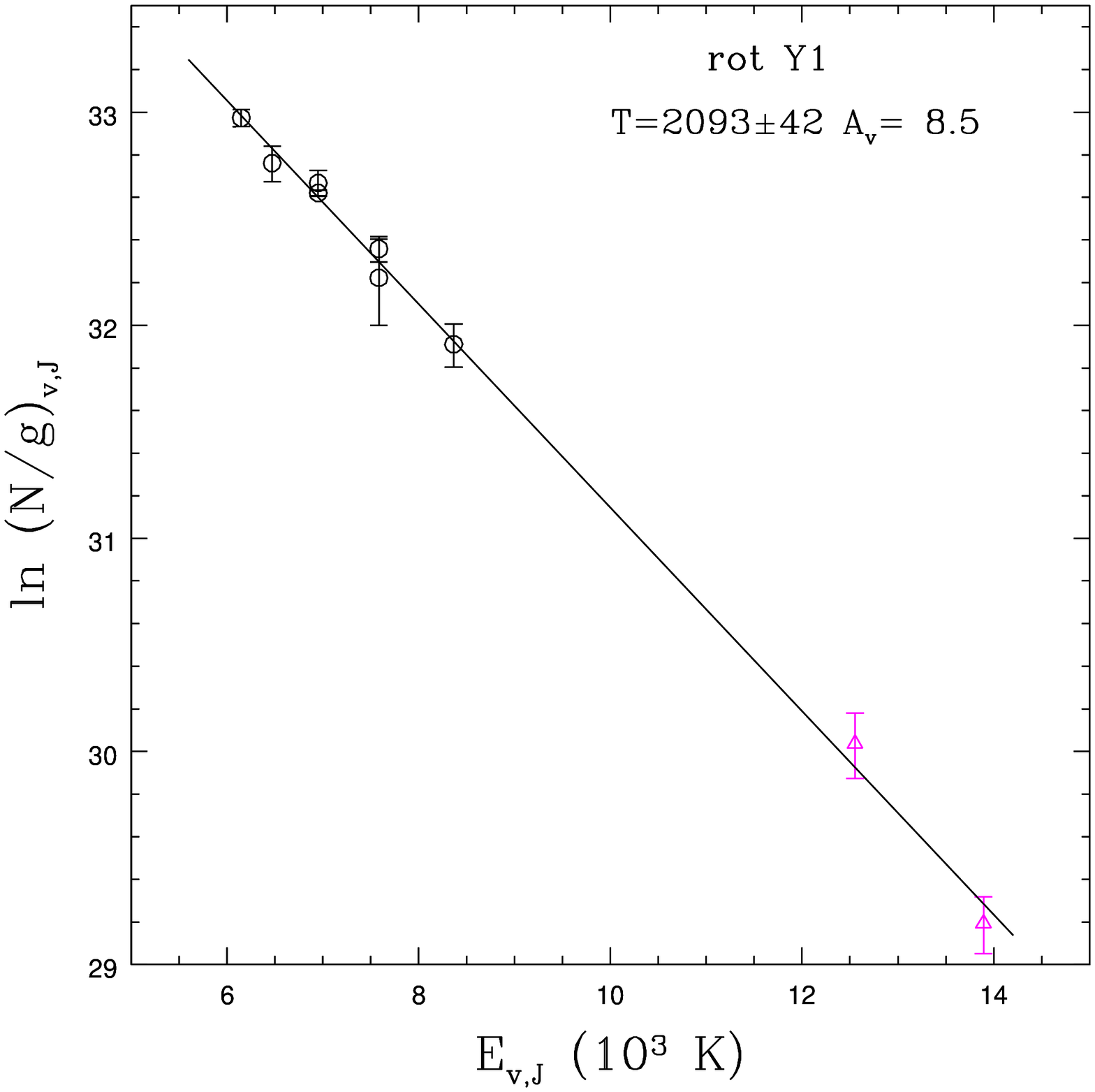}
\centering
\caption{\label{knot_y1} As Fig.\,\ref{fig:rot_z2} for knot Y1}
\end{figure}

\begin{figure}[!h]
 
\includegraphics[totalheight=7.2cm]{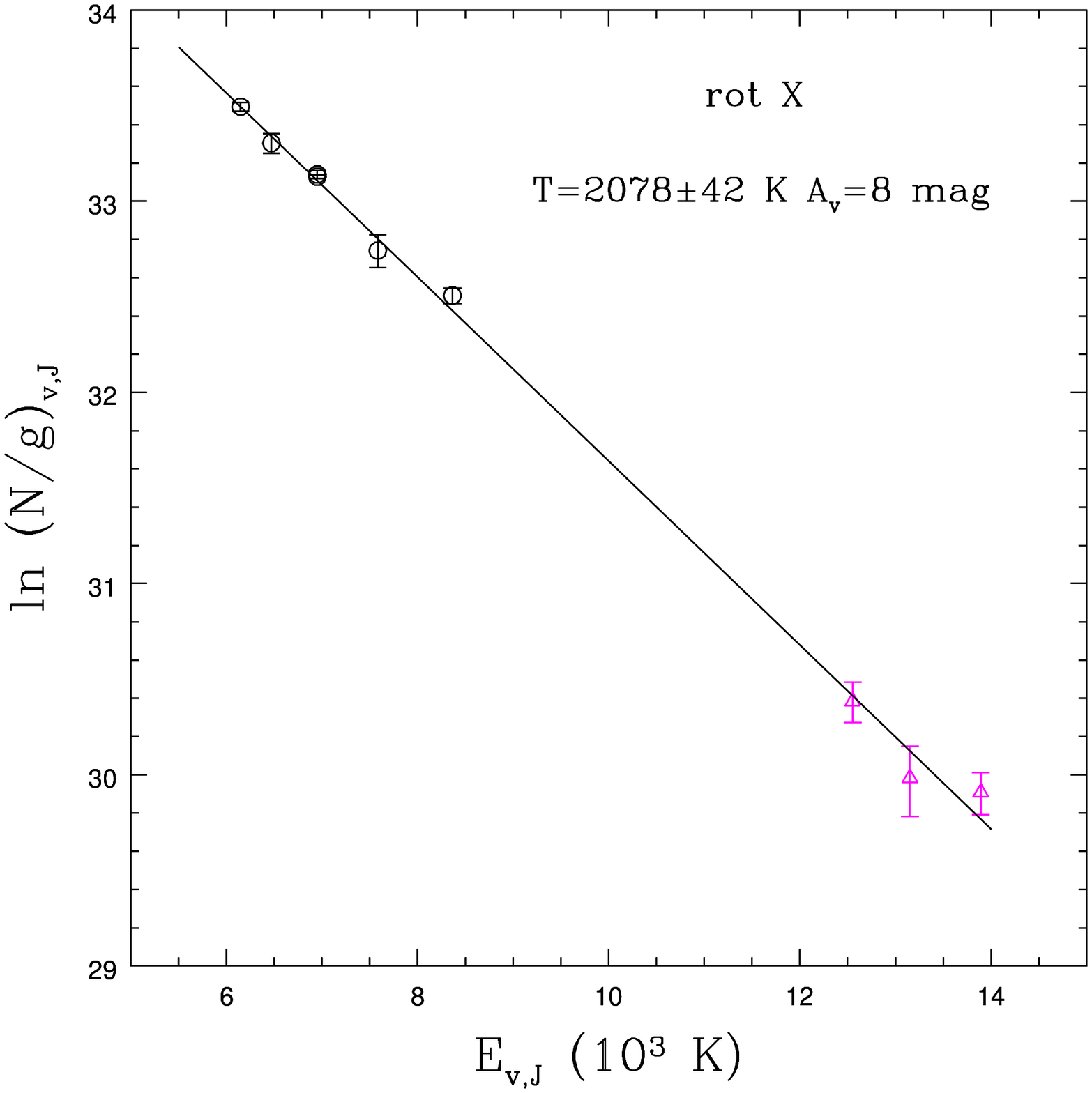}
\centering
\caption{\label{knot_Y}As Fig.\,\ref{fig:rot_z2} for knot X}
\end{figure}

\end{appendix}

\end{document}